\documentclass{jpsj3}
\usepackage{txfonts}

\title{Negative-Temperature State Formed and Interactions Inverted by Symmetric Monocycle Optical Pulse Excitation}

\author{Kenji Yonemitsu$^{1,2}$\thanks{E-mail: kxy@phys.chuo-u.ac.jp} and Keita Nishioka$^3$}
\inst{$^1$Department of Physics, Chuo University, Bunkyo, Tokyo 112-8551, Japan \\
$^2$JST, CREST, Chiyoda, Tokyo 102-0076, Japan \\
$^3$Math and Science Education Research Center, Kanazawa Institute of Technology, Nonoichi, Ishikawa 921-8501, Japan} 

\abst{
The excited state dynamics of correlated electron and electron-phonon systems triggered by an oscillating electric-field pulse of large amplitude are theoretically investigated. A ``negative-temperature'' state and inversion of electron-electron and electron-phonon interactions are induced even by a symmetric monocycle pulse. This fact is numerically demonstrated, using the exact diagonalization method, in a band-insulator phase of one-dimensional three-quarter-filled strongly dimerized extended Peierls-Hubbard and Holstein models. When the total-energy increment is maximized as a function of the electric field amplitude, the occupancy of the bonding and antibonding orbitals is inverted to produce a negative-temperature state. Around this state, the dependences of time-averaged electron-electron and electron-phonon correlation functions on interaction parameters are opposite to those in the ground state.
}

\begin{document}
\maketitle

\section{Introduction}

Photoinduced changes in electronic properties of correlated electron and electron-phonon systems have been investigated extensively so far.\cite{koshigono_jpsj06,yonemitsu_pr08} Development in experimental techniques has allowed for progressively shortening  optical pulses and consequently improving time resolutions. At the same time, the field amplitude that can be applied to materials has significantly increased for different excitation energies. This fact enhances the possibility of controlling the electronic properties. 

As for events that occur with an oscillating electric field of large amplitude, dynamical localization is known in the continuous-wave case.\cite{dunlap_prb86,grossmann_prl91} When the frequency of the field is comparable to or higher than that of electron transfers determined by the transfer integral, the slow component of the electron dynamics is governed by the time average of the transfer integral with the Peierls-phase factor.\cite{kayanuma_pra94} The time average for a sinusoidally oscillating field gives an effective transfer integral modulated by the zeroth-order Bessel function whose argument is proportional to the ratio of the amplitude to the frequency of the field. Thus, the effective transfer integral can vanish to cause dynamical localization or can even change from positive (negative) to negative (positive). 

Applying such a strong field to interacting electron systems, Tsuji {\it et al.} have shown that the sign inversion of the transfer integral and the consequent inversion of the band structure, a ``negative-temperature'' state, are regarded as the sign inversion of the interaction in the half-filled Hubbard model, using the dynamical mean-field theory, by which a superconducting state is possibly induced.\cite{tsuji_prl11} In addition, they have shown that such a repulsion-to-attraction conversion is realized not only by continuous waves but also by half-cycle pulses and asymmetric monocycle pulses.\cite{tsuji_prb12} In the limit of the vanishing pulse width, the irradiation causes a sudden change in the Peierls phase of the transfer integral, and the dynamical phase shift is equivalent to the crystal momentum shift. In this limit, however, a symmetric monocycle pulse causes nothing. Indeed, the repulsion-to-attraction conversion is hardly realized by symmetric monocycle pulses with various trial time profiles of finite pulse widths.\cite{tsuji_prb12} 

The effective transfer integral modulated by the application of a continuous wave and that by the application of a pulse are not always the same. It depends on the system and how the system is excited. The quasi-two-dimensional metal complex Et$_2$Me$_2$Sb[Pd(dmit)$_2$]$_2$ (dmit = 1,3-dithiol-2-thione-4,5-dithiolate) is known to show a photoinduced transition from a charge-ordered-insulator phase to a Mott-insulator phase.\cite{ishikawa_prb09,nishioka_jpsj13a,nishioka_jpsj13b} In this particular case, it has been demonstrated that continuous-wave lasers modulate the effective transfer integrals between the dimers, and pulsed lasers modulate that within the dimer, both through the zeroth-order Bessel function.\cite{nishioka_jpsj14} On the basis of this fact, the dynamical localization is transiently expected for pulsed lasers if the field amplitude is sufficiently large. Indeed, optical freezing of charge motion has been found in an organic conductor, $\alpha$-(BEDT-TTF)$_2$I$_3$ (BEDT-TTF = bis[ethylenedithio]-tetrathiafulvalene).\cite{ishikawa_ncomms14} Although its mechanism is not theoretically understood yet, it is regarded as due to collaboration with the dynamical localization and intermolecular charge correlations. 

The realization of a negative-temperature state or modulation of transfer integrals and interaction parameters requires an oscillating electric field of large amplitude. In general, it is much larger than what is needed for some phase transition to be simply induced. In any case, the fact that transfer integrals are effectively modulated by pulsed lasers is encouraging because the field amplitude can be increased by shortening the pulse width. In the actual case of $\alpha$-(BEDT-TTF)$_2$I$_3$ above,\cite{ishikawa_ncomms14} the dynamical phase shift is known to be much smaller than $\pi$. On the basis of these findings, we study the effect of symmetric monocycle electric-field pulses whose time integrals are zero, i.e., whose dynamical phase shifts are zero. It was expected to be difficult to form a negative-temperature state or to invert interaction parameters by such symmetric pulses.\cite{tsuji_prb12} The model systems we treat are similar to that used in Ref.~\citen{nishioka_jpsj14}, but they are simplified to use the exact diagonalization method: one-dimensional three-quarter-filled strongly dimerized extended Peierls-Hubbard and Holstein models. Their band-insulator phases consist of almost half-filled dimers alternating with almost completely filled dimers. Charge transfers take place between these dimers, but the optical gap is determined by the electronic transition within the almost half-filled dimer, as in Et$_2$Me$_2$Sb[Pd(dmit)$_2$]$_2$. 

In this paper, we show that, even by a symmetric monocycle electric-field pulse, a negative-temperature state is produced, and interactions are effectively inverted when we interpret the dependences of the time-averaged correlation functions on the interactions. Namely, the stronger on-site repulsion causes the larger space-time-averaged double occupancy, the stronger nearest-neighbor repulsion causes the larger averaged nearest-neighbor charge-density correlation, and so on. Such a negative-temperature state appears when the total-energy increment is maximized as a function of the electric field amplitude. 

\section{Strongly Dimerized Models with Charge-Ordered Ground States}

\begin{figure}
\includegraphics[height=9cm]{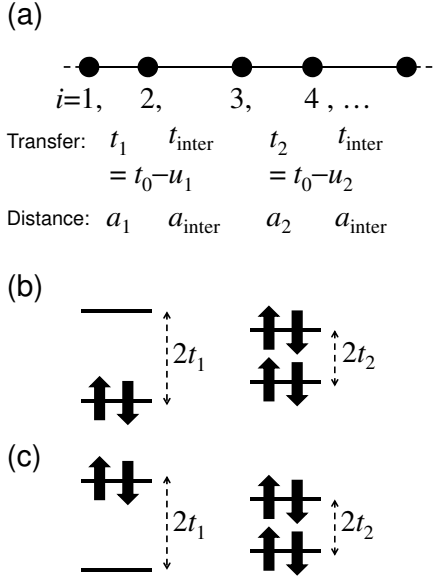}
\caption{
(a) Schematic representation of transfer integrals and distances between neighboring sites.
(b) Half-filled dimer whose bonding orbitals are occupied and completely filled dimer in ground state.
(c) Half-filled dimer whose antibonding orbitals are occupied and completely filled dimer in negative-temperature state. 
\label{fig:negativeT}}
\end{figure}

We study the electron-phonon dynamics after strong photoexcitation in a one-dimensional three-quarter-filled strongly dimerized model with on-site and nearest-neighbor repulsions and with the Peierls and Holstein types of electron-phonon couplings, 
\begin{eqnarray}
H & = & -\sum_{j=1}^{N/2} (t_0-u_j) \sum_\sigma \left( 
c_{2j-1,\sigma}^\dagger c_{2j,\sigma} + c_{2j,\sigma}^\dagger c_{2j-1,\sigma}
\right) \nonumber \\ & & 
-\sum_{j=1}^{N/2} t_{\rm inter} \sum_\sigma \left( 
c_{2j,\sigma}^\dagger c_{2j+1,\sigma} + c_{2j+1,\sigma}^\dagger c_{2j,\sigma}
\right) \nonumber \\ & & 
+ U \sum_{i=1}^{N} n_{i,\uparrow} n_{i,\downarrow} 
+ V \sum_{i=1}^{N} n_{i} n_{i+1} 
+ \sum_{j=1}^{N/2} \left( 
\frac{u_j^2}{2S} + \frac{\dot{u}_j^2}{2S\omega_P^2} 
\right) \nonumber \\ & & 
+ g \sum_{i=1}^{N} (b_i+b_i^\dagger) \left( n_i - \frac{3}{2} \right)
+ \omega_H \sum_{i=1}^{N} b_i^\dagger b_i
\;, \label{eq:hamiltonian}
\end{eqnarray}
where $ c_{i,\sigma}^\dagger $ creates an electron with spin $ \sigma $ at site $ i $, $ n_{i,\sigma} = c_{i,\sigma}^\dagger c_{i,\sigma} $, and $ n_i = \sum_\sigma n_{i,\sigma} $. The transfer integral within the $ j $-th dimer (i.e., between sites $ 2j-1 $ and $ 2j $) is modulated by the lattice displacement $ u_j $: $ t_0-u_j $. The transfer integral between the dimers is denoted by $ t_{\rm inter} $. They are shown schematically in Fig.~\ref{fig:negativeT}(a).
The parameter $ U $ represents the onsite repulsion strength, and $ V $ represents the nearest-neighbor repulsion strength. The lattice displacement $ u_j $ has a bare frequency $ \omega_P $ and is coupled with the bond density between sites $ 2j-1 $ and $ 2j $ with strength $ S $. It is noted that $ u $'s have a dimension of energy. The operator $ b_i^\dagger $ creates a phonon, which has a bare frequency $ \omega_H $ and is coupled with the charge density relative to its average value at site $ i $ with strength $ g $. 

The periodic boundary condition is imposed. For $ g $=0, we use $ N $=16 sites. For $ g\neq 0 $, we use $ N $=8 sites and limit the number of phonons up to a maximum of two on every site. The number of electrons is 24 for $ N $=16 and 12 for $ N $=8 at three-quarter filling. The electron-quantum-phonon system is treated by the exact diagonalization method. The lattice system is treated classically. 
This is because the lattice vibration is caused by the intermolecular restoring force and consequently is slow compared with other dynamics. Meanwhile, the molecular vibration is caused by the intramolecular restoring force and its energy is often comparable to those of the intermolecular charge transfers. Its quantum nature is important and essential for the polaron effect. 
Before the photoexcitation, the system is in the lowest-energy state so that the lattice displacements are determined by 
\begin{equation}
u_j = -S \sum_\sigma \left( 
\langle c_{2j-1,\sigma}^\dagger c_{2j,\sigma} \rangle + 
\langle c_{2j,\sigma}^\dagger c_{2j-1,\sigma} \rangle 
\right) 
\;. \label{eq:initial_displacement}
\end{equation}
They spontaneously alternate large and weak distortions, to which $ u_1 $ and $ u_2 $ are assigned [Fig.~\ref{fig:negativeT}(a)]. The electron-quantum-phonon ground state is in the band-insulator phase, which is schematically shown in Fig.~\ref{fig:negativeT}(b) with $ t_1 > t_2 > 0 $: it consists of strongly ($ u_1 < 0 $, large $ \mid u_1 \mid $) distorted dimers with the bonding orbitals occupied and weakly ($ u_2 < 0 $, small $ \mid u_2 \mid $) distorted dimers with both the bonding and antibonding orbitals occupied. 

Photoexcitation is introduced through the Peierls phase 
\begin{equation}
c_{i,\sigma}^\dagger c_{j,\sigma} \rightarrow
\exp \left[
\frac{ie}{\hbar c} r_{ij} A(t)
\right] c_{i,\sigma}^\dagger c_{j,\sigma}
\;, \label{eq:photo_excitation}
\end{equation}
where $ r_{ij}=r_j-r_i $ and $ r_i $ being the location of the $ i $-th site, which is substituted into Eq.~(\ref{eq:hamiltonian}). For symmetric monocycle electric-field pulses, we use the time-dependent vector potential, 
\begin{equation}
A(t) = \frac{cF}{\omega} \left[ \cos (\omega t)-1 \right] 
\theta (t) \theta \left( \frac{2\pi}{\omega}-t \right)
\;, \label{eq:monocycle_pulse}
\end{equation}
where $ F $ is the field amplitude and the central frequency $ \omega $ is chosen to be nearly resonant with the optical gap. The time-dependent Schr\"odinger equation for the exact many-electron-quantum-phonon wave function is numerically solved by expanding the exponential evolution operator with a time slice $ dt $=0.02 to the 15th order and by checking the conservation of the norm. \cite{yonemitsu_prb09,matsubara_prb14} We use the leapfrog method to solve the classical equations for the lattice displacements, 
\begin{equation}
\frac{1}{S\omega_P^2} \frac{d^2 u_j}{dt^2} = F_j(\left\{ u \right\}) 
\;, \label{eq:lattice_motion}
\end{equation}
where the force $ F_j(\left\{ u \right\}) $ is obtained with the aid of the Hellmann-Feynman theorem: 
\begin{equation}
F_j(\left\{ u \right\}) =
-\sum_\sigma \left[ 
e^{\frac{ie}{\hbar c}r_{2j-1,2j}A(t)} 
\langle c_{2j-1,\sigma}^\dagger c_{2j,\sigma} \rangle + 
e^{\frac{ie}{\hbar c}r_{2j,2j-1}A(t)} 
\langle c_{2j,\sigma}^\dagger c_{2j-1,\sigma} \rangle 
\right] -\frac{u_j}{S} 
\;. \label{eq:force}
\end{equation}

For the bare transfer integrals, we use $ t_0 $=1 and $ t_{\rm inter} $=0.1, which correspond to an extrinsically and strongly dimerized system. The interaction parameters $ U $, $ V $, $ g $, and $ S $ are varied. The Holstein coupling strength $ g $ is chosen to be so weak that the number of phonons is always much smaller than one on every site, $ \langle \Psi (t) \mid b_i^\dagger b_i \mid \Psi (t) \rangle \ll 1 $. The bare phonon frequency $ \omega_H $, corresponding to a molecular vibration frequency, is set to be $ \omega_H=0.2 $. The bare phonon frequency $ \omega_P $, corresponding to an intermolecular oscillation frequency, is set to be $ \omega_P=0.05 $ if only $ S $ is nonzero among the couplings $ U $, $ V $, $ g $, and $ S $ (i.e., in the Peierls model) and $ \omega_P=0 $ with $ S=0.1 $ otherwise (i.e., in the extended Hubbard and Holstein models with frozen lattice displacements). The intermolecular distances $ r_{ij} $ are fixed for simplicity and so chosen that the Bessel functions $ J_0(er_{ij}F/\hbar\omega) $ modulating the transfers between sites $ i $ and $ j=i+1 $ ($ i $=1, 3 for the two nonequivalent intradimer transfers and $ i $=2, 4 for the interdimer transfers) are easily distinguishable: $ a_1 \equiv r_{12} = a $, $ a_2 \equiv r_{34} = 1.2a $, and $ a_{\rm inter} \equiv r_{23} = r_{45} = 1.6a $. For $ S=0.1 $, $ t_1=t_0-u_1 \simeq 1.2 $ and $ t_2=t_0-u_2 \simeq 1.0 $ [Fig.~\ref{fig:negativeT}(a)]. The difference between $ t_1 $ and $ t_2 $ becomes smaller for larger $ U $. 

\begin{figure}
\includegraphics[height=5.6cm]{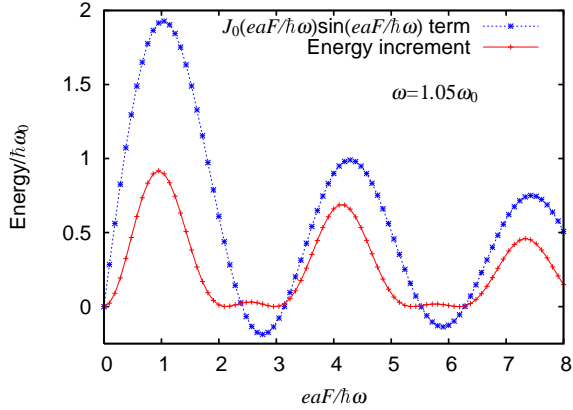}
\caption{(Color online) 
Energy increment and $ J_0(eaF/\hbar\omega)\sin(eaF/\hbar\omega) $ term for monocycle nearly resonant ($ \omega=1.05\omega_0 $) excitation of one-electron two-site tight-binding model.\cite{nishioka_jpsj14} 
\label{fig:nishioka_A13A11}}
\end{figure}

As pointed out previously in Appendix of Ref.~\citen{nishioka_jpsj14} for the two-site tight-binding model and for a resonantly excited strongly dimerized model, the total-energy increment $ \Delta E_{\rm tot} $ is roughly proportional to $ J_0(eaF/\hbar\omega) \sin(eaF/\hbar\omega)  $ although $ \Delta E_{\rm tot} $ is nonnegative. 
For reference, we show again, in Fig.~\ref{fig:nishioka_A13A11}, the energy increment [Eq.~(A$\cdot$11) in Ref.~\citen{nishioka_jpsj14}] and the term proportional to $ J_0(eaF/\hbar\omega)\sin(eaF/\hbar\omega) $ [Eq.~(A$\cdot$13) in Ref.~\citen{nishioka_jpsj14}] for the monocycle nearly resonant ($ \omega=1.05\omega_0 $ with the bonding-to-antibonding transition energy $ \hbar \omega_0=2t $) excitation of the one-electron two-site tight-binding model with transfer integral $ t $ (upper panel of Fig.~A$\cdot$3 in Ref.~\citen{nishioka_jpsj14}). 
The maximum of $ J_0(eaF/\hbar\omega)\sin(eaF/\hbar\omega) $ appears around $ eaF/\hbar\omega  $=1, while the maximum of the energy increment appears at $ eaF/\hbar\omega  $ slightly smaller than 1. 

\section{States after Symmetric Monocycle Pulse Excitation}

\begin{figure}
\includegraphics[height=11.2cm]{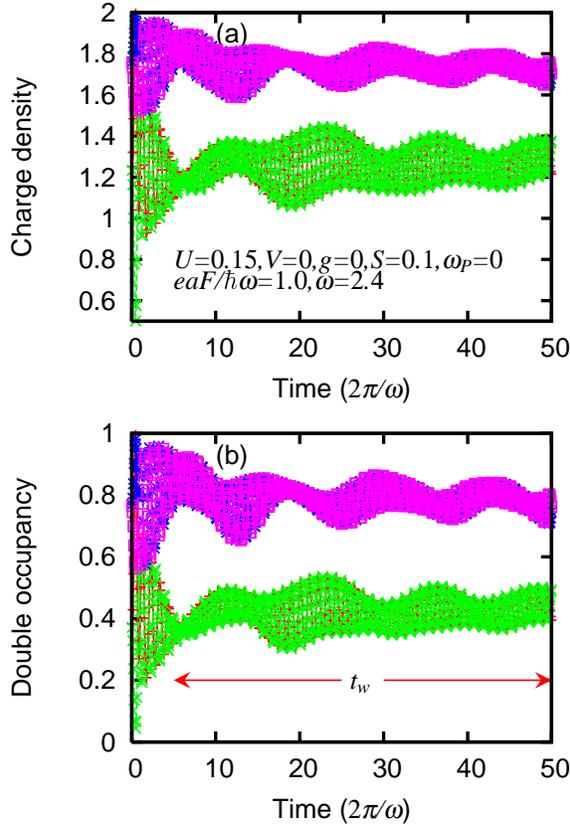}
\caption{(Color online) 
(a) Charge density $ \langle \Psi (t) \mid n_i \mid \Psi (t) \rangle $ and (b) double occupancy $ \langle \Psi (t) \mid n_{i,\uparrow} n_{i,\downarrow} \mid \Psi (t) \rangle $, at four nonequivalent sites during and after monocycle pulse excitation with $ eaF/\hbar\omega=1.0 $ and $ \omega=2.4 $. The model parameters are $ U=0.15 $, $ V=0 $, $ g=0 $, $ S=0.1 $, and $ \omega_P=0 $. The time averages below are taken for the time region indicated by $ t_w $. 
\label{fig:t_evol_U0p15f2p4_nd}}
\end{figure}

As a typical case, the time evolution of the charge density $ \langle \Psi (t) \mid n_i \mid \Psi (t) \rangle $ and that of the double occupancy $ \langle \Psi (t) \mid n_{i,\uparrow} n_{i,\downarrow} \mid \Psi (t) \rangle $ are shown in Figs.~\ref{fig:t_evol_U0p15f2p4_nd}(a) and \ref{fig:t_evol_U0p15f2p4_nd}(b), respectively, for $ U=0.15 $, $ V=0 $, $ g=0 $, and $ 0 < t < 50 T $. 
The charge density is, initially and a short time after the pulse excitation $ t > 5T $, almost 1, 1, 2, 2 for sites $ i $=1, 2, 3, 4, respectively [see also Fig.~\ref{fig:negativeT}(a), denoted formally by the (1,1,2,2) configuration hereafter]. Because the present system is in the band-insulator phase, the time evolution of the double occupancy is quite similar to that of the charge density. The time-averaged double occupancy below is calculated by 
\begin{equation}
\langle \langle n_{i,\uparrow} n_{i,\downarrow} \rangle \rangle =
\frac{1}{t_w} \int_{t_s}^{t_s+t_w} 
\langle \Psi (t) \mid n_{i,\uparrow} n_{i,\downarrow} \mid \Psi (t) \rangle dt
\;, \label{eq:time_average}
\end{equation}
with $ t_s=5 T $, $ t_w=45 T $, and $ T=2\pi/\omega $ being the period of the oscillating electric field. The time domain of integration is indicated by the two-headed arrow. The other time-averaged quantities are calculated likewise.

\subsection{Effect of on-site repulsion $ U $}

\begin{figure}
\includegraphics[height=11.2cm]{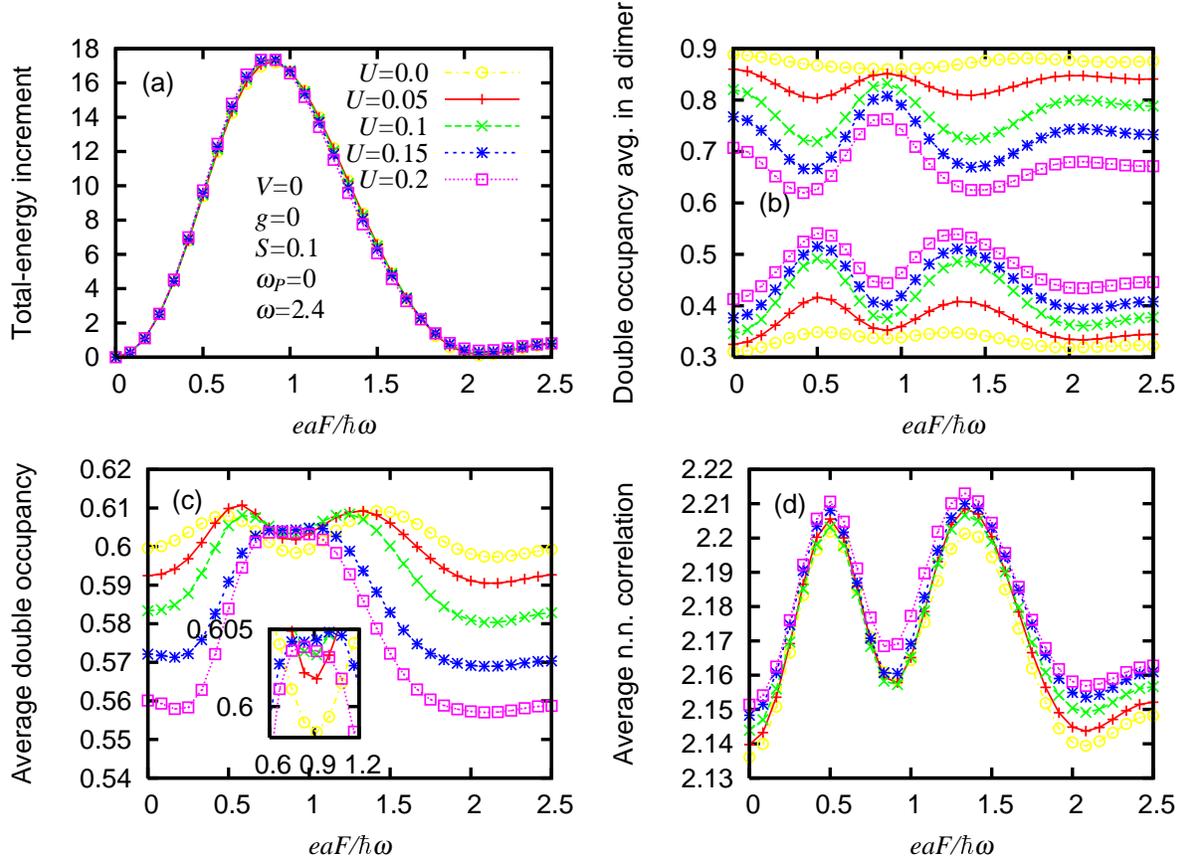}
\caption{(Color online) 
(a) Total-energy increment $ \Delta E_{\rm tot} $, (b) time-averaged double occupancy averaged over strongly distorted dimer sites $ (\langle \langle n_{1,\uparrow} n_{1,\downarrow} \rangle \rangle + \langle \langle n_{2,\uparrow} n_{2,\downarrow} \rangle \rangle)/2 $ and that averaged over weakly distorted dimer sites $ (\langle \langle n_{3,\uparrow} n_{3,\downarrow} \rangle \rangle + \langle \langle n_{4,\uparrow} n_{4,\downarrow} \rangle \rangle)/2 $, (c) average double occupancy $ \frac{1}{N} \sum_{i=1}^{N} \langle \langle n_{i,\uparrow} n_{i,\downarrow} \rangle \rangle $, and (d) average nearest-neighbor charge-density correlation $ \frac{1}{N} \sum_{i=1}^{N} \langle \langle n_i n_{i+1} \rangle \rangle $, as a function of $ eaF/\hbar\omega $, for different values of $ U $. The other parameters are $ V=0 $, $ g=0 $, $ S=0.1 $, $ \omega_P=0 $, and $ \omega=2.4 $. In (c), the vicinity of $ eaF/\hbar\omega $=0.9 is enlarged in the inset. 
\label{fig:Fw_d_nn_U_2p5}}
\end{figure}

Time-averaged quantities for $ V=0 $ and $ g=0 $ with different values of $ U $ and fixed $ \omega $=2.4 are shown in Fig.~\ref{fig:Fw_d_nn_U_2p5} in the range of $ 0< eaF/\hbar\omega <2.5 $. 
The total-energy increment $ \Delta E_{\rm tot} $ is almost independent of $ U $ in the range of $ U < 0.23 $ [Fig.~\ref{fig:Fw_d_nn_U_2p5}(a)]. The maximum appears around $ eaF/\hbar\omega $=0.9. As is clear from the comparison between Figs.~\ref{fig:nishioka_A13A11} and \ref{fig:Fw_d_nn_U_2p5}(a), it is governed by the Bessel function $ J_0(eaF/\hbar\omega) $ modulating the transfer between sites 1 and 2 within the strongly distorted dimer: the system is excited by the electronic transitions within the strongly distorted dimers. The maximum value is indeed close to the energy increment expected when all electrons in the strongly distorted dimers transit from the bonding orbitals to the antibonding orbitals [Fig.~\ref{fig:negativeT}(c)]: $ 2 t_1 \times 2 \times $(16 sites/4 sites)$ \simeq 2.3 \times 2 \times 4 \simeq 18$. 

As $ eaF/\hbar\omega  $ increases from zero to about 0.5, the time-averaged charge density $ \langle \langle n_{i} \rangle \rangle $ changes from the (1,1,2,2) configuration toward a uniform distribution, especially for large $ U $, so that the charge order is weakened and the time-averaged double occupancy $ \langle \langle n_{i,\uparrow} n_{i,\downarrow} \rangle \rangle $ also changes toward a uniform distribution [Fig.~\ref{fig:Fw_d_nn_U_2p5}(b)]. However, around $ eaF/\hbar\omega $=0.9, where $ \Delta E_{\rm tot} $ reaches the maximum, the configurations of $ \langle \langle n_{i} \rangle \rangle $ and $ \langle \langle n_{i,\uparrow} n_{i,\downarrow} \rangle \rangle $ almost return to those before the photoexcitation, which correspond to $ eaF/\hbar\omega $=0. This indicates that, in the corresponding state, all electrons in the strongly distorted dimers indeed transit from the bonding orbitals to the antibonding orbitals [Fig.~\ref{fig:negativeT}(c)] and almost no electrons are transferred between dimers. If only these intradimer transitions were realized, the site-diagonal quantities such as $ \langle \langle n_{i} \rangle \rangle $, $ \langle \langle n_{i,\uparrow} n_{i,\downarrow} \rangle \rangle $, and the nearest-neighbor charge-density correlation $ \langle \langle n_i n_{i+1} \rangle \rangle $ would be unchanged. This very state is called a negative-temperature state. 

In the ground state and weakly excited states of small $ eaF/\hbar\omega $, as $ U $ increases, the average double occupancy decreases [Fig.~\ref{fig:Fw_d_nn_U_2p5}(c)]. Around $ eaF/\hbar\omega $=0.9, however, the average double occupancy for $ U > 0 $ is larger than that of the noninteracting ($ U $=0) ground state ($ eaF/\hbar\omega $=0) and increases as $ U $ increases. This behavior with increasing $ U $ corresponds to more attractive on-site interactions for equilibrium: it appears as if the negative-temperature state has an attractive on-site interaction and its strength increases with $ U $ when the time-averaged correlation is interpreted in terms of the correlation in equilibrium. Near the maximum of $ \Delta E_{\rm tot} $, the energy supplied by the photoexcitation goes slightly to the $ U $ term in Eq.~(\ref{eq:hamiltonian}) to enhance the average double occupancy, as discussed later in Sect.~\ref{subsec:flow}. 
For small $ eaF/\hbar\omega $, as $ U $ increases, the average nearest-neighbor charge-density correlation increases [Fig.~\ref{fig:Fw_d_nn_U_2p5}(d)]. Around $ eaF/\hbar\omega $=0.9, this correlation is larger than that of the ground state, slightly decreases as $ U $ varies from $ U $=0 to $ U $=0.1, and then increases as $ U $ varies from $ U $=0.1 to $ U $=0.2. This fact cannot be interpreted in terms of an effective on-site interaction for equilibrium. 

\begin{figure}
\includegraphics[height=11.2cm]{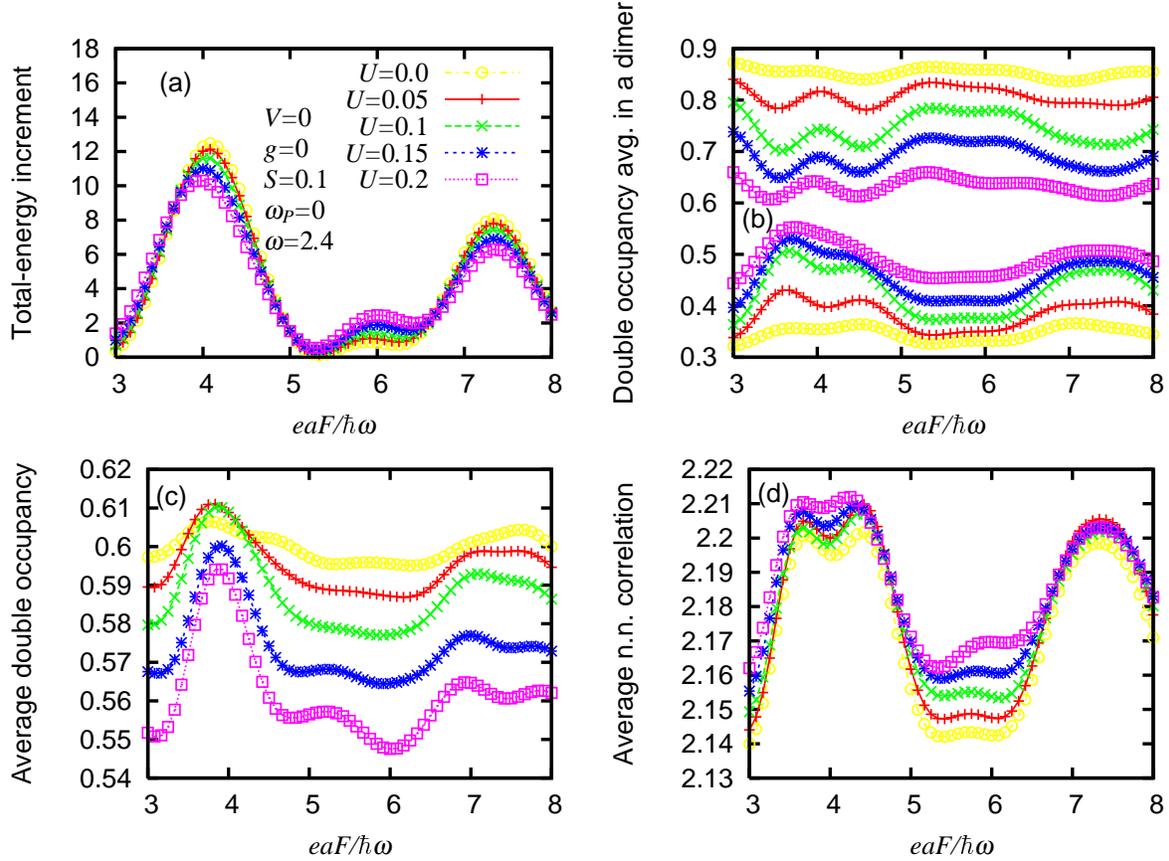}
\caption{(Color online) 
Same quantities as in Fig.~\ref{fig:Fw_d_nn_U_2p5}, with the same ordinate axes and for $ 3< eaF/\hbar\omega <8 $. 
\label{fig:Fw_d_nn_U_3_8}}
\end{figure}

The same quantities as in Fig.~\ref{fig:Fw_d_nn_U_2p5} are shown in Fig.~\ref{fig:Fw_d_nn_U_3_8} for larger field amplitudes, $ 3< eaF/\hbar\omega <8 $. 
For comparison, the same ordinate axes are used. Local maxima of the total-energy increment $ \Delta E_{\rm tot} $ appear around $ eaF/\hbar\omega $=4 and $ eaF/\hbar\omega $=7.3 [Fig.~\ref{fig:Fw_d_nn_U_3_8}(a)]. Their values depart from the maximum value with increasing $ eaF/\hbar\omega $. Around $ eaF/\hbar\omega $=4, the weakened charge order recovers to a small extent [Fig.~\ref{fig:Fw_d_nn_U_3_8}(b), compare it with Fig.~\ref{fig:Fw_d_nn_U_2p5}(b)]. Around $ eaF/\hbar\omega $=7.3, the weakened charge order does not recover at all. The realized state differs from the typical negative-temperature state at $ eaF/\hbar\omega $=0.9 because the electrons are transferred from the weakly distorted dimers to the strongly distorted dimers. Thus, the charge order is weaker and $ \Delta E_{\rm tot} $ is smaller than the corresponding quantities at $ eaF/\hbar\omega $=0.9. Around $ eaF/\hbar\omega $=4 and $ eaF/\hbar\omega $=7.3, as $ U $ increases, the average double occupancy decreases except for $ eaF/\hbar\omega $=4 with $ 0 < U < 0.1 $ [Fig.~\ref{fig:Fw_d_nn_U_3_8}(c)]: the effective on-site attraction exists for $ eaF/\hbar\omega $=4 and $ 0 < U < 0.1 $ only. The average nearest-neighbor charge-density correlation indeed increases as $ U $ increases even around them [Fig.~\ref{fig:Fw_d_nn_U_3_8}(d)]. 

\subsection{Effect of nearest-neighbor repulsion $ V $}

\begin{figure}
\includegraphics[height=11.2cm]{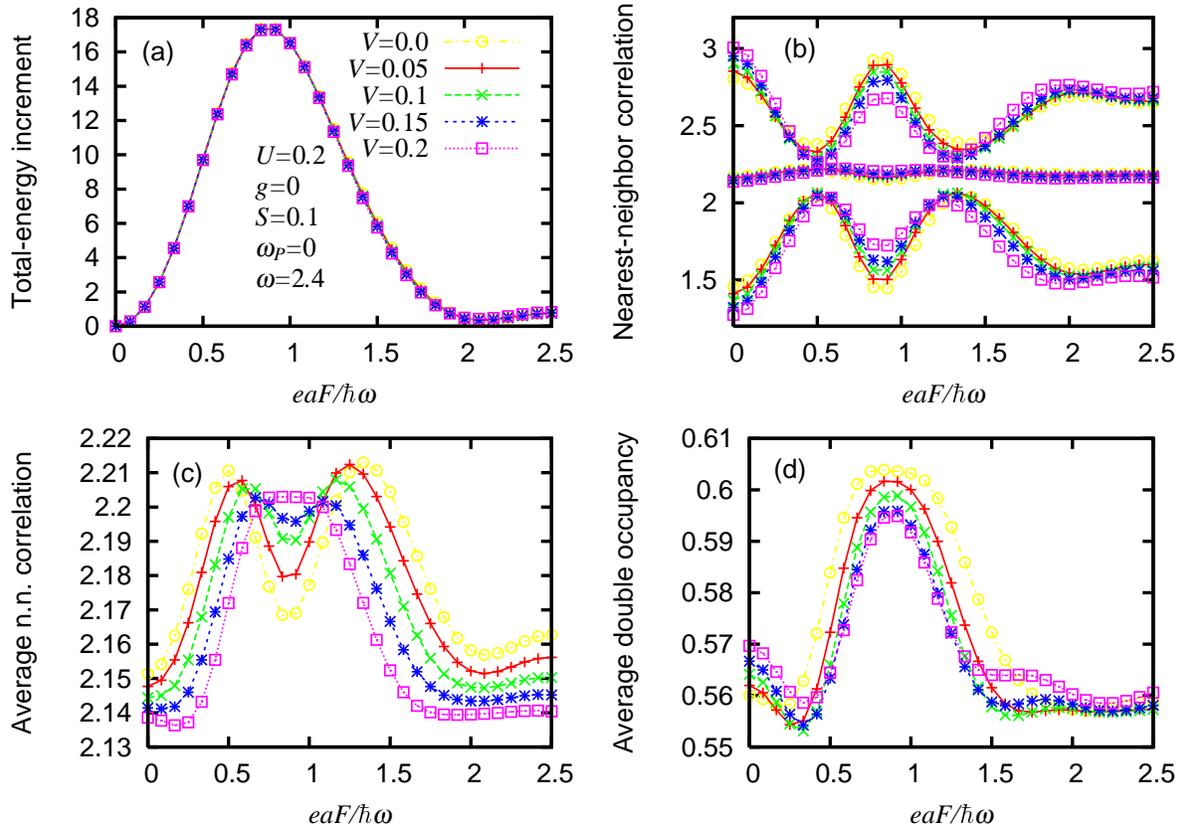}
\caption{(Color online) 
(a) Total-energy increment $ \Delta E_{\rm tot} $, (b) time-averaged nearest-neighbor charge-density correlation $ \langle \langle n_i n_{i+1} \rangle \rangle $ at interdimer ($ i=2 $) and two nonequivalent intradimer ($ i=1,3 $) bonds, (c) average nearest-neighbor charge-density correlation $ \frac{1}{N} \sum_{i=1}^{N} \langle \langle n_i n_{i+1} \rangle \rangle $, and (d) average double occupancy $ \frac{1}{N} \sum_{i=1}^{N} \langle \langle n_{i,\uparrow} n_{i,\downarrow} \rangle \rangle $, as a function of $ eaF/\hbar\omega $, for different values of $ V $. The other parameters are $ U=0.2 $, $ g=0 $, $ S=0.1 $, $ \omega_P=0 $, and $ \omega=2.4 $. 
\label{fig:Fw_nn_d_UmV_2p5}}
\end{figure}

Time-averaged quantities for $ U=0.2 $ and $ g=0 $ with different values of $ V $ and fixed $ \omega=2.4 $ are shown for $ 0 < eaF/\hbar\omega < 2.5 $ in Fig.~\ref{fig:Fw_nn_d_UmV_2p5}. 
The total-energy increment $ \Delta E_{\rm tot} $ is almost independent of $ V $ [Fig.~\ref{fig:Fw_nn_d_UmV_2p5}(a)]. The maximum appears around $ eaF/\hbar\omega $=0.9. The maximum value is again close to the energy increment expected when Fig.~\ref{fig:negativeT}(c) is realized. 
As $ eaF/\hbar\omega $ increases from zero to about 0.5, $ \langle \langle n_{i} \rangle \rangle $ changes from the (1,1,2,2) configuration toward a uniform distribution so that the charge order is weakened and $ \langle \langle n_i n_{i+1} \rangle \rangle $ also changes toward a uniform distribution [Fig.~\ref{fig:Fw_nn_d_UmV_2p5}(b)]. However, around $ eaF/\hbar\omega $=0.9, the configurations of $ \langle \langle n_{i} \rangle \rangle $ and $ \langle \langle n_i n_{i+1} \rangle \rangle $ almost return to those before the photoexcitation. Namely, the negative-temperature state [Fig.~\ref{fig:negativeT}(c)] is realized. 

In the ground state and weakly excited states of small $ eaF/\hbar\omega $, as $ V $ increases, the average nearest-neighbor charge-density correlation decreases [Fig.~\ref{fig:Fw_nn_d_UmV_2p5}(c)] and the average double occupancy increases [Fig.~\ref{fig:Fw_nn_d_UmV_2p5}(d)]. Around $ eaF/\hbar\omega $=0.9, the average nearest-neighbor charge-density correlation is larger than that of the ground state ($ eaF/\hbar\omega $=0) for $ V $=0 and increases as $ V $ increases. At the same place, the average double occupancy decreases as $ V $ increases. These behaviors with increasing $ V $ are consistent with more attractive nearest-neighbor interactions for equilibrium. Near the maximum of $ \Delta E_{\rm tot} $, the supplied energy goes to the $ V $ term in Eq.~(\ref{eq:hamiltonian}) to enhance the average nearest-neighbor charge-density correlation, as discussed later in Sect.~\ref{subsec:flow}. 

A typical negative-temperature state does not appear at $ eaF/\hbar\omega $=4 or $ eaF/\hbar\omega $=7.3 (not shown), where a local maximum of $ \Delta E_{\rm tot} $ appears. Around $ eaF/\hbar\omega $=4, the weakened charge order recovers to a small extent. Around $ eaF/\hbar\omega $=7.3, the weakened charge order does not recover at all. In both cases, as $ V $ increases, the average nearest-neighbor charge-density correlation decreases except for $ eaF/\hbar\omega $=4 with $ 0 < V < 0.05 $: the effective nearest-neighbor attraction exists for $ eaF/\hbar\omega $=4 and $ 0 < V < 0.05 $ only. How these states deviate from the typical negative-temperature state is similar to that shown previously in Fig.~\ref{fig:Fw_d_nn_U_3_8}. 

\subsection{Effect of Holstein electron-phonon coupling $ g $}

\begin{figure}
\includegraphics[height=11.2cm]{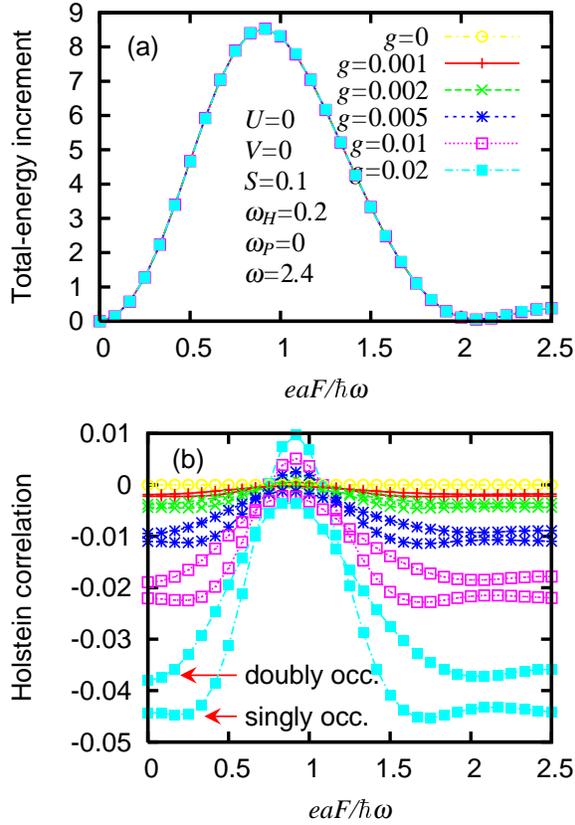}
\caption{(Color online) 
(a) Total-energy increment $ \Delta E_{\rm tot} $ and (b) time-averaged Holstein electron-phonon correlation averaged over almost singly occupied sites $ (\langle \langle (b_1+b_1^\dagger)(n_1-3/2) \rangle \rangle + \langle \langle (b_2+b_2^\dagger)(n_2-3/2) \rangle \rangle)/2 $ and that averaged over almost doubly occupied sites $ (\langle \langle (b_3+b_3^\dagger)(n_3-3/2) \rangle \rangle + \langle \langle (b_4+b_4^\dagger)(n_4-3/2) \rangle \rangle)/2 $, as a function of $ eaF/\hbar\omega $, for different values of $ g $. The other parameters are $ U=0 $, $ V=0 $, $ S=0.1 $, $ \omega_H=0.2 $, $ \omega_P=0 $, and $ \omega=2.4 $. 
\label{fig:Fw_hlst_G_2p5}}
\end{figure}

The total-energy increment $ \Delta E_{\rm tot} $ and the time-averaged Holstein electron-phonon correlation $ \langle \langle (b_i+b_i^\dagger)(n_i-3/2) \rangle \rangle $ for $ U=0 $, $ V=0 $, and $ \omega_H=0.2 $ with different values of $ g $ and fixed $ \omega=2.4 $ are shown for $ 0 < eaF/\hbar\omega < 2.5 $ in Figs.~\ref{fig:Fw_hlst_G_2p5}(a) and \ref{fig:Fw_hlst_G_2p5}(b), respectively.  
The maximum of $ \Delta E_{\rm tot} $ appears around $ eaF/\hbar\omega $=0.9 again. The maximum value is close to the energy increment expected when Fig.~\ref{fig:negativeT}(c) is realized: $ 2 t_1 \times 2 \times $(8 sites/4 sites)$ \simeq 2.3 \times 2 \times 2 \simeq 9$. 

In the ground state and weakly excited states of small $ eaF/\hbar\omega $, the correlations $ \langle \langle (b_i+b_i^\dagger)(n_i-3/2) \rangle \rangle $ are negative and their magnitudes increase as $ g $ increases [Fig.~\ref{fig:Fw_hlst_G_2p5}(b)]. 
Note that, in equilibrium, the magnitude of the correlation $ \langle (b_i+b_i^\dagger)(n_i-3/2) \rangle $ is larger at the almost singly occupied sites even though the intramolecular displacement $ \langle b_i+b_i^\dagger \rangle $ at the almost singly occupied sites and that at the almost doubly occupied sites have opposite signs and the same magnitude, as $ \langle n_i-3/2 \rangle $ at the almost singly occupied sites ($ \simeq -0.5 $) and that at the almost doubly occupied sites ($ \simeq +0.5 $). 
As $ eaF/\hbar\omega $ increases from zero to about 0.7, the correlations are weakened so that the magnitudes of $ \langle \langle (b_i+b_i^\dagger)(n_i-3/2) \rangle \rangle $ become small. Around $ eaF/\hbar\omega $=0.9, $ \langle \langle (b_i+b_i^\dagger)(n_i-3/2) \rangle \rangle $ at the almost singly occupied sites is positive and increases as $ g $ increases. On the other hand, $ \langle \langle (b_i+b_i^\dagger)(n_i-3/2) \rangle \rangle $ at the almost doubly occupied sites remains negative, but its magnitude is quite small around $ eaF/\hbar\omega $=0.9. If $ \langle \langle (b_i+b_i^\dagger)(n_i-3/2) \rangle \rangle $ are spatially averaged over all sites, the average correlation is positive and increases with $ g $ around $ eaF/\hbar\omega $=0.9. This behavior corresponds to inverted electron-phonon couplings (from positive $ g $ to negative $ g $) for equilibrium. The energy supplied by the photoexcitation goes slightly to the $ g $ term in Eq.~(\ref{eq:hamiltonian}) to invert the average Holstein correlation, as discussed later in Sect.~\ref{subsec:flow}.  

This fact is consistent with the finding in Ref.~\citen{kawakami_prl10}: at the early stage of the photoinduced melting of charge order in $\alpha$-(BEDT-TTF)$_2$I$_3$, the oscillation of the correlated electrons between the neighboring molecules interferes destructively with intramolecular vibrations. Their coupling is theoretically treated by the Holstein term.\cite{kawakami_prl10} When a sufficient amount of energy goes to the $ g $ term, the phase of the electronic oscillation relative to the phase of the intramolecular vibration becomes opposite to that in equilibrium. It is because, in the ground state, the relative phase is determined in such a way that the interaction energy is lowered. This change in the relative phase is observed as the destructive interference. 

A typical negative-temperature state does not appear at $ eaF/\hbar\omega $=4 or $ eaF/\hbar\omega $=7.3 (not shown), where a local maximum of $ \Delta E_{\rm tot} $ appears. 
Around $ eaF/\hbar\omega $=4, the magnitudes of $ \langle \langle (b_i+b_i^\dagger)(n_i-3/2) \rangle \rangle $ are small, but they remain negative. Around $ eaF/\hbar\omega $=7.3, the magnitudes of $ \langle \langle (b_i+b_i^\dagger)(n_i-3/2) \rangle \rangle $ become slightly smaller than the neighborhood, but they are not so small compared with those around $ eaF/\hbar\omega $=4. Namely, the realized state departs from any state with an inverted interaction. 

\subsection{Effect of Peierls electron-lattice coupling $ S $}

\begin{figure}
\includegraphics[height=11.2cm]{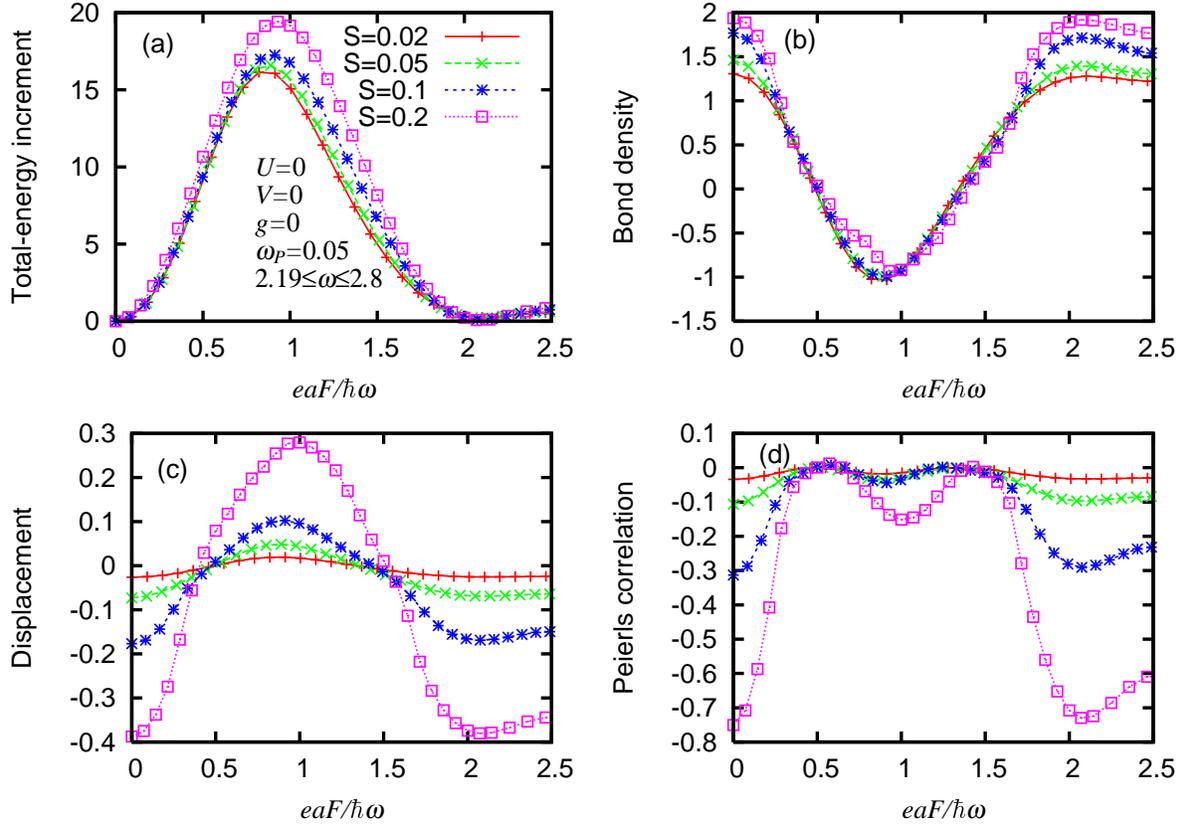}
\caption{(Color online) 
(a) Total-energy increment $ \Delta E_{\rm tot} $, (b) time-averaged bond density $ \langle \langle \sum_\sigma \left( c_{1,\sigma}^\dagger c_{2,\sigma} + c_{2,\sigma}^\dagger c_{1,\sigma} \right) \rangle \rangle $, (c) time-averaged lattice displacement $ \langle \langle u_1 \rangle \rangle $, and (d) time-averaged Peierls electron-lattice correlation $ \langle \langle u_1 \sum_\sigma \left( c_{1,\sigma}^\dagger c_{2,\sigma} + c_{2,\sigma}^\dagger c_{1,\sigma} \right) \rangle \rangle $, as a function of $ eaF/\hbar\omega $, for different values of $ S $. The other parameters are $ U=0 $, $ V=0 $, $ g=0 $, and $ \omega_P=0.05 $. The central frequency is set at $ \omega=2.19 $ for $ S=0.02 $, $ \omega=2.25 $ for $ S=0.05 $, $ \omega=2.41 $ for $ S=0.1 $, and $ \omega=2.8 $ for $ S=0.2 $. 
\label{fig:Fw_bnd_frc_prl_S_2p5}}
\end{figure}

Time-averaged quantities for $ U=0 $, $ V=0 $, $ g=0 $, and $ \omega_P=0.05 $ with different values of $ S $ and $ \omega $ are shown for $ 0 < eaF/\hbar\omega < 2.5 $ in Fig.~\ref{fig:Fw_bnd_frc_prl_S_2p5}. 
The optical gap sensitively depends on $ S $, so that the central frequency $ \omega $, which is chosen to be resonant with the optical gap, is varied: $ \omega=2.19 $ for $ S=0.02 $, $ \omega=2.25 $ for $ S=0.05 $, $ \omega=2.41 $ for $ S=0.1 $, and $ \omega=2.8 $ for $ S=0.2 $. As $ S $ increases, the energy difference between the bonding and antibonding orbitals in the strongly distorted dimer increases. Thus, the maximum of $ \Delta E_{\rm tot} $, which appears around $ eaF/\hbar\omega $=0.9, also increases with $ S $ [Fig.~\ref{fig:Fw_bnd_frc_prl_S_2p5}(a)]. The maximum value is again close to the energy increment expected when Fig.~\ref{fig:negativeT}(c) is realized. 

As $ eaF/\hbar\omega $ increases, the time-averaged bond density between the almost singly occupied sites $ \langle \langle \sum_\sigma \left( c_{1,\sigma}^\dagger c_{2,\sigma} + c_{2,\sigma}^\dagger c_{1,\sigma} \right) \rangle \rangle $ decreases and then becomes negative [Fig.~\ref{fig:Fw_bnd_frc_prl_S_2p5}(b)]. Around $ eaF/\hbar\omega $=0.9, its magnitude shows a local maximum. Although its magnitude is smaller than that in the ground state, this corresponds to the negative-temperature state shown in Fig.~\ref{fig:negativeT}(c). Around $ eaF/\hbar\omega $=2, where $ \Delta E_{\rm tot} $ almost vanishes, the time-averaged bond density almost takes the value taken by the ground state, i.e., the time-averaged bond density almost remains the initial value before the photoexcitation. In the ground state and weakly excited states of small $ eaF/\hbar\omega $, the time-averaged lattice displacement between the almost singly occupied sites $ \langle \langle u_1 \rangle \rangle $ is negative and its magnitude increases as $ S $ increases [Fig.~\ref{fig:Fw_bnd_frc_prl_S_2p5}(c)]. As $ eaF/\hbar\omega $ increases, $ \langle \langle u_1 \rangle \rangle $ increases and then becomes positive. Around $ eaF/\hbar\omega $=0.9, $ \langle \langle u_1 \rangle \rangle $ shows a local maximum, which increases with $ S $. The magnitude of $ \langle \langle u_1 \rangle \rangle $ here is close to, but smaller than, that in the ground state. 

Actually, around $ eaF/\hbar\omega $=2, all time-averaged quantities almost take the values taken by the ground state in all cases of Figs.~ \ref{fig:Fw_d_nn_U_2p5}, \ref{fig:Fw_nn_d_UmV_2p5}, \ref{fig:Fw_hlst_G_2p5}, and \ref{fig:Fw_bnd_frc_prl_S_2p5}. This situation is similar to the time behavior of the double occupancy in the cases $ eaF/\hbar\omega $=2 and $ eaF/\hbar\omega $=2.5 of Ref.~\citen{tsuji_prl11} after the sudden application of a continuous-wave field to the half-filled Hubbard model in the dynamical mean-field theory. Although the double occupancy is shown for a short time in Ref.~\citen{tsuji_prl11}, its time average would be equal to its initial value for some value of $ eaF/\hbar\omega $ slightly smaller than 2.5. Note that the dynamical localization is expected for $ eaF/\hbar\omega $=2.4 in the continuous-wave case. 

In the ground state and weakly excited states of small $ eaF/\hbar\omega $, the time-averaged Peierls electron-lattice correlation between the almost singly occupied sites $ \langle \langle u_1 \sum_\sigma \left( c_{1,\sigma}^\dagger c_{2,\sigma} + c_{2,\sigma}^\dagger c_{1,\sigma} \right) \rangle \rangle $ is negative and its magnitude increases as $ S $ increases [Fig.~\ref{fig:Fw_bnd_frc_prl_S_2p5}(d)]. As $ eaF/\hbar\omega $ increases from zero to about 0.3, the correlation is weakened so that the magnitude of $ \langle \langle u_1 \sum_\sigma \left( c_{1,\sigma}^\dagger c_{2,\sigma} + c_{2,\sigma}^\dagger c_{1,\sigma} \right) \rangle \rangle $ becomes small. Around $ eaF/\hbar\omega $=0.5 and $ eaF/\hbar\omega $=1.3, $ \langle \langle u_1 \sum_\sigma \left( c_{1,\sigma}^\dagger c_{2,\sigma} + c_{2,\sigma}^\dagger c_{1,\sigma} \right) \rangle \rangle $ is quite small but positive for $ S > 0.05 $ and increases as $ S $ increases. The energy used to invert the Peierls correlation is discussed below in Sect.~\ref{subsec:flow}. The field amplitudes $ eaF/\hbar\omega $=0.5 and $ eaF/\hbar\omega $=1.3 correspond to those that weaken the charge order to the largest degree in Figs.~ \ref{fig:Fw_d_nn_U_2p5}(b) and \ref{fig:Fw_nn_d_UmV_2p5}(b). Namely, the electrons are transferred most extensively between dimers. This is the reason why $ \langle \langle u_1 \sum_\sigma \left( c_{1,\sigma}^\dagger c_{2,\sigma} + c_{2,\sigma}^\dagger c_{1,\sigma} \right) \rangle \rangle $ is quite small. 
If both the time-averaged bond density and $ \langle \langle u_1 \rangle \rangle $ invert their signs, their correlation should not invert its sign. This happens around $ eaF/\hbar\omega $=0.9. Here also, the magnitude of their correlation increases as $ S $ increases. This corresponds to the negative-temperature state. Namely, the negative-temperature state appears when $ \Delta E_{\rm tot} $ reaches its maximum ($ eaF/\hbar\omega $=0.9), while the inverted electron-lattice coupling [$(t_0-u_j)$ to $(t_0+u_j)$ in Eq.~(\ref{eq:hamiltonian})] appears in different field amplitudes ($ eaF/\hbar\omega $=0.5 and $ eaF/\hbar\omega $=1.3). 

\subsection{Energy flows and inverted interactions \label{subsec:flow}}

So far, we mainly show whether the correlation functions merely increase or decrease as a function of the coupling strength or the field amplitude. In this section, we focus on those photoinduced states whose correlation functions behave as if the interactions are inverted. We show how the inverted correlations are related to changes in the total energy and in the interaction energy. 

\begin{figure}
\includegraphics[height=11.2cm]{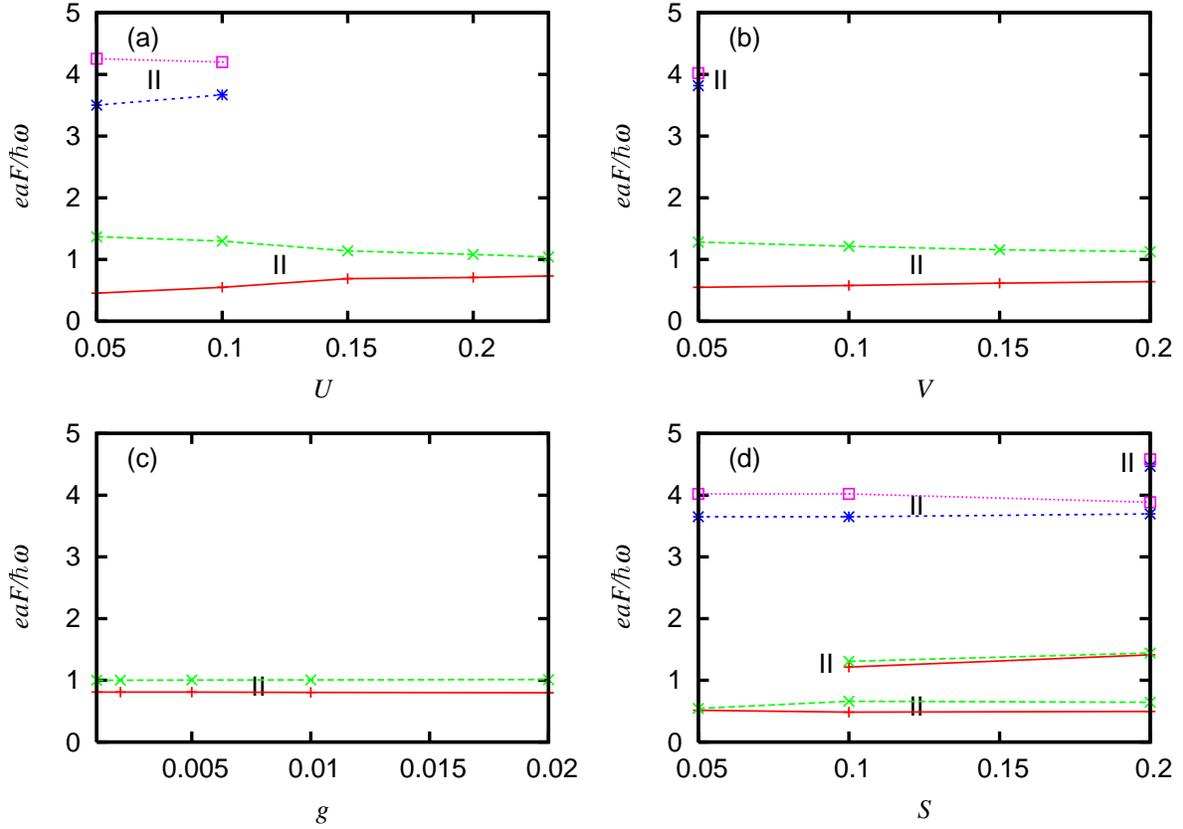}
\caption{(Color online) 
Regions of inverted interactions, denoted by `II', in planes spanned by $ eaF/\hbar\omega $ and 
(a) $ U $ ($ V $=0, $ g $=0, $ S $=0.1, and $ \omega_P $=0), 
(b) $ V $ ($ U $=0.2, $ g $=0, $ S $=0.1, and $ \omega_P $=0), 
(c) $ g $ ($ U $=0, $ V $=0, $ S $=0.1, $ \omega_H $=0.2, and $ \omega_P $=0), and (d) $ S $ ($ U $=0, $ V $=0, $ g $=0, and $ \omega_P $=0.05), 
from data shown in Figs.~ \ref{fig:Fw_d_nn_U_2p5}(c)--\ref{fig:Fw_d_nn_U_3_8}(c), \ref{fig:Fw_nn_d_UmV_2p5}(c), \ref{fig:Fw_hlst_G_2p5}(b), and \ref{fig:Fw_bnd_frc_prl_S_2p5}(d), respectively, for resonant excitations. 
\label{fig:UVGS_DeltaCorr}}
\end{figure}

The enhanced average double occupancy is defined by 
\begin{equation}
\frac{1}{N} \sum_{i=1}^{N} \Delta_U 
\langle \langle n_{i,\uparrow} n_{i,\downarrow} \rangle \rangle = 
\frac{1}{N} \sum_{i=1}^{N} \left(
\langle \langle n_{i,\uparrow} n_{i,\downarrow} \rangle \rangle \mid_{U>0} - 
\langle \langle n_{i,\uparrow} n_{i,\downarrow} \rangle \rangle \mid_{U=0} 
\right)
\;,
\end{equation}
for fixed $ eaF/\hbar\omega $. The regions where the enhanced average double occupancy is positive are shown as the regions of the inverted interaction in Fig.~\ref{fig:UVGS_DeltaCorr}(a). 
They appear around the local maxima of $ \Delta E_{\rm tot} $. For $ eaF/\hbar\omega $=0.92 and $ U $=0.15, the enhanced average double occupancy is 6$\times$10$^{-3}$ in Fig.~\ref{fig:Fw_d_nn_U_2p5}(c). This value is rather small compared with the $ U $ dependence of the average double occupancy in equilibrium. The energy flowed to enhance the average double occupancy $ \Delta_U E_U $ is defined as 
\begin{equation}
\Delta_U E_U = U \sum_{i=1}^{N} \Delta_U 
\langle \langle n_{i,\uparrow} n_{i,\downarrow} \rangle \rangle 
\;.
\end{equation}
In the case above, $ \Delta_U E_U $ is 1.4$\times$10$^{-2}$. We further define $ \Delta_U \Delta E_{\rm tot} $ by 
\begin{equation}
\Delta_U \Delta E_{\rm tot} = 
\Delta E_{\rm tot} \mid_{U>0} - \Delta E_{\rm tot} \mid_{U=0}
\;,
\end{equation}
for fixed $ eaF/\hbar\omega $. In the case above, $ \Delta_U \Delta E_{\rm tot} $ is 1.8$\times$10$^{-1}$, while $ \Delta E_{\rm tot} $ is 17. The ratio $ \Delta_U E_U / \Delta_U \Delta E_{\rm tot} $ is then 7.8$\times$10$^{-2}$, while the ratio $ \Delta_U E_U / \Delta E_{\rm tot} $ is 8.3$\times$10$^{-4}$. The increased average double occupancy is defined by 
\begin{equation}
\frac{1}{N} \sum_{i=1}^{N} \Delta 
\langle \langle n_{i,\uparrow} n_{i,\downarrow} \rangle \rangle = 
\frac{1}{N} \sum_{i=1}^{N} \left( 
\langle \langle n_{i,\uparrow} n_{i,\downarrow} \rangle \rangle \mid_{F>0} - 
\langle \langle n_{i,\uparrow} n_{i,\downarrow} \rangle \rangle \mid_{F=0} 
\right) 
\;,
\end{equation}
for fixed $ U $. Note that this quantity is negative for $ U $=0 and positive for $ U \geq $0.05 in Fig.~\ref{fig:Fw_d_nn_U_2p5}(c). Using this quantity, the interaction-energy increment $ \Delta E_U $ is evaluated as 
\begin{equation}
\Delta E_U = U \sum_{i=1}^{N} \Delta 
\langle \langle n_{i,\uparrow} n_{i,\downarrow} \rangle \rangle
\;.
\end{equation}
For $ eaF/\hbar\omega $=0.92 and $ U $=0.15, the increased average double occupancy is 3.2$\times$10$^{-2}$ in Fig.~\ref{fig:Fw_d_nn_U_2p5}(c), and $ \Delta E_U $ is 7.7$\times$10$^{-2}$, so that $ \Delta E_U / \Delta E_{\rm tot} $ is 4.4$\times$10$^{-3}$. 

The enhanced average nearest-neighbor charge-density correlation is defined by 
\begin{equation}
\frac{1}{N} \sum_{i=1}^{N} \Delta_V 
\langle \langle n_{i} n_{i+1} \rangle \rangle = 
\frac{1}{N} \sum_{i=1}^{N} \left(
\langle \langle n_{i} n_{i+1} \rangle \rangle \mid_{V>0} -
\langle \langle n_{i} n_{i+1} \rangle \rangle \mid_{V=0} 
\right)
\;,
\end{equation}
for fixed $ eaF/\hbar\omega $. The regions where the enhanced average nearest-neighbor charge-density correlation is positive are shown as the regions of the inverted interaction in Fig.~\ref{fig:UVGS_DeltaCorr}(b). They appear around the local maxima of $ \Delta E_{\rm tot} $. For $ eaF/\hbar\omega $=0.92 and $ V $=0.2, the enhanced average nearest-neighbor charge-density correlation is 3.4$\times$10$^{-2}$ in Fig.~\ref{fig:Fw_nn_d_UmV_2p5}(c). This value is rather large compared with the $ V $ dependence of the average nearest-neighbor charge-density correlation in equilibrium. The energy flowed to enhance the average nearest-neighbor charge-density correlation is evaluated as 
\begin{equation}
\Delta_V E_V = V \sum_{i=1}^{N} \Delta_V 
\langle \langle n_{i} n_{i+1} \rangle \rangle
\;.
\end{equation}
In the case above, $ \Delta_V E_V $ is 1.1$\times$10$^{-1}$. The quantity $ \Delta_V \Delta E_{\rm tot} $, defined by 
\begin{equation}
\Delta_V \Delta E_{\rm tot} = 
\Delta E_{\rm tot} \mid_{V>0} - \Delta E_{\rm tot} \mid_{V=0}
\;,
\end{equation}
is 5.5$\times$10$^{-2}$, while $ \Delta E_{\rm tot} $ is 17. The ratio $ \Delta_V E_V / \Delta_V \Delta E_{\rm tot} $ is then about 2, while the ratio $ \Delta_V E_V / \Delta E_{\rm tot} $ is 6.3$\times$10$^{-3}$. The fact that $ \Delta_V E_V / \Delta_V \Delta E_{\rm tot} $ is larger than 1 is realized by the relation $ \Delta_V E_U < 0 $, as shown in Fig.~\ref{fig:Fw_nn_d_UmV_2p5}(d): the large enhancement of the average nearest-neighbor charge-density correlation is realized by the reduction of the average double occupancy in the system with competing interactions $ U $ and $ V $. The increased average nearest-neighbor charge-density correlation, which is defined by 
\begin{equation}
\frac{1}{N} \sum_{i=1}^{N} \Delta 
\langle \langle n_{i} n_{i+1} \rangle \rangle =
\frac{1}{N} \sum_{i=1}^{N} \left(
\langle \langle n_{i} n_{i+1} \rangle \rangle \mid_{F>0} - 
\langle \langle n_{i} n_{i+1} \rangle \rangle \mid_{F=0}
\right)
\;,
\end{equation}
for fixed $ V $, is 6.5$\times$10$^{-2}$ in Fig.~\ref{fig:Fw_nn_d_UmV_2p5}(c), and the interaction-energy increment $ \Delta E_V $, 
\begin{equation}
\Delta E_V = V \sum_{i=1}^{N} \Delta 
\langle \langle n_{i} n_{i+1} \rangle \rangle 
\;,
\end{equation}
is 2.1$\times$10$^{-1}$, so that $ \Delta E_V / \Delta E_{\rm tot} $ is 1.2$\times$10$^{-2}$ in the case above. 

The enhanced average Holstein correlation is given by the average of the quantities for the almost singly occupied and doubly occupied sites shown in Fig.~\ref{fig:Fw_hlst_G_2p5}(b) because it is zero for $ g $=0. The region where the enhanced average Holstein correlation is positive is shown as the region of the inverted interaction in Fig.~\ref{fig:UVGS_DeltaCorr}(c). It appears only around the maximum of $ \Delta E_{\rm tot} $. Because we used small $ g $ for the technical reason mentioned before, every energy scale is small. 

The enhanced average Peierls correlation is given by the quantity shown in Fig.~\ref{fig:Fw_bnd_frc_prl_S_2p5}(d) because it is zero for $ S $=0. The regions where the enhanced average Peierls correlation is positive are shown as the regions of the inverted interaction in Fig.~\ref{fig:UVGS_DeltaCorr}(d). They appear on both sides of the local maxima of $ \Delta E_{\rm tot} $. For $ eaF/\hbar\omega $=0.57 and $ S $=0.2, this quantity is 1.2$\times$10$^{-2}$ in Fig.~\ref{fig:Fw_bnd_frc_prl_S_2p5}(d). The energy flowed to enhance the average Peierls correlation $ \Delta_S E_S $, 
\begin{equation}
\Delta_S E_S = \sum_{j=1}^{N/2} \langle \langle u_j \sum_\sigma \left( 
c_{2j-1,\sigma}^\dagger c_{2j,\sigma} + c_{2j,\sigma}^\dagger c_{2j-1,\sigma} 
\right) \rangle \rangle 
\;,
\end{equation}
is 5$\times$10$^{-2}$. The quantity $ \Delta_S \Delta E_{\rm tot} $ is defined as before, and  $ \Delta_S \Delta E_{\rm tot} $ is 1.8, while $ \Delta E_{\rm tot} $ is 13. The ratio $ \Delta_S E_S / \Delta_S \Delta E_{\rm tot} $ is 2.7$\times$10$^{-2}$, while the ratio $ \Delta_S E_S / \Delta E_{\rm tot} $ is 3.6$\times$10$^{-3}$. The increased average Peierls correlation is 7.6$\times$10$^{-1}$ in Fig.~\ref{fig:Fw_bnd_frc_prl_S_2p5}(d), and the interaction-energy increment $ \Delta E_S $, 
\begin{equation}
\Delta E_S = \sum_{j=1}^{N/2} \left[
\langle \langle u_j \sum_\sigma \left( 
c_{2j-1,\sigma}^\dagger c_{2j,\sigma} + c_{2j,\sigma}^\dagger c_{2j-1,\sigma} 
\right) \rangle \rangle \mid_{F>0} - 
\langle \langle u_j \sum_\sigma \left( 
c_{2j-1,\sigma}^\dagger c_{2j,\sigma} + c_{2j,\sigma}^\dagger c_{2j-1,\sigma} 
\right) \rangle \rangle \mid_{F=0}
\right]
\;,
\end{equation}
is about 3.0, so that $ \Delta E_S / \Delta E_{\rm tot} $ is 0.23. This ratio is relatively large because the ground state is stabilized by the Peierls electron-lattice coupling in the first place. 

Comparing Figs.~\ref{fig:UVGS_DeltaCorr}(c) with \ref{fig:UVGS_DeltaCorr}(d), we notice that the region of the inverted interaction is wider for $ S $ than for $ g $. When we see the local maxima of $ \Delta E_{\rm tot} $, there is indeed a difference between them. For $ g \neq 0 $, with increasing $ eaF/\hbar\omega $, the deviation from the typical negative-temperature state becomes large. For $ S \neq 0 $ but $ g=0 $, with increasing $ eaF/\hbar\omega $, the deviation does not seem to become so large at least for $ eaF/\hbar\omega < 10 $. In the latter, the disordering effect seems small. If we increased $ \omega_P $ and treated $ u_j $ quantum-mechanically, the disordering effect might become large to suppress the inversion of the Peierls electron-phonon coupling for large $ eaF/\hbar\omega $. 

In all cases above, for the interaction parameter $ X $ ($ X $=$ U $, $ V $, $ g $, or $ S $), the average correlation enhanced by the interaction $ X $ is smaller than the average correlation increased by the field $ F $. Even in the system with competing interactions $ U $ and $ V $, the ratio of the average nearest-neighbor charge-density correlation enhanced by $ V $ to that increased by $ F $ is about 1 to 2. When the interaction-induced change is compared with the field-induced change for the energy ratio, we find $ \Delta_X E_X / \Delta_X \Delta E_{\rm tot} > \Delta E_X / \Delta E_{\rm tot} $ for $ X $=$ U $ and $ V $, while $ \Delta_X E_X / \Delta_X \Delta E_{\rm tot} < \Delta E_X / \Delta E_{\rm tot} $ for $ X $=$ S $. In any case, $ \Delta_X E_X / \Delta E_{\rm tot} $ is less than $ 10^{-2} $ and very small. The efficiency of the inversion of interactions is quite low. 

\section{Discussion}

In general, the formation of a negative-temperature state requires suppressing the disordering effect (i.e., thermalization) on increasing the total energy. The present model has a simple structure. The strongly distorted dimer is regarded as a type of two-level system, which is a typical system to achieve a negative-temperature state. The weakly distorted dimers are then regarded as the environment, to which the energy flow from the system is allowed. The interdimer charge transfer is a type of relaxation process. The structure of the present model may be advantageous to forming a negative-temperature state in the sense that the thermalization is suppressed by separating the energy scale of the interdimer electron transfer processes and that of the intradimer process. 

In real substances, the network of electron transfers is not always simple as treated in this study. For instance, in the model previously used for Et$_2$Me$_2$Sb[Pd(dmit)$_2$]$_2$,\cite{nishioka_jpsj13a} the weakly distorted dimer has an unoccupied molecular orbital whose energy is not far from those of the antibonding orbitals in the strongly distorted dimer. In such a case, even under optimal conditions (such as $ eaF/\hbar\omega $=0.9 in the present study), the electrons can be easily transferred between dimers so that the negative-temperature state, if realized, can be short-lived, depending on the transfer integral between the corresponding molecular orbitals. We expect that, if the photoinduced melting of charge order (i.e., a high-temperature state) is easily achieved, the negative-temperature state would hardly be achieved, and vice versa. 

\section{Conclusions}

Electron-phonon dynamics in the band-insulator phase of one-dimensional three-quarter-filled strongly dimerized extended Peierls-Hubbard and Holstein models after the application of a symmetric monocycle electric-field pulse with central frequency resonant with the optical gap are calculated using the exact diagonalization method. From the field-amplitude and interaction-parameter dependences of time-averaged correlation functions, we discuss the negative-temperature state and the inversion of interactions. The total-energy increment is maximized when the energy gain of an electron transferred to the neighboring site $ eaF $ under the electric field $ F $ is comparable to the photoexcitation energy $ \hbar\omega $ ($ eaF/\hbar\omega $=0.9). The maximum value is close to the energy increment expected when all electrons in the strongly distorted dimers transit from the bonding orbitals to the antibonding orbitals, i.e., the energy difference between the negative-temperature state and the ground state. In this state, the site-diagonal quantities such as the time-averaged charge density, double occupancy, and nearest-neighbor charge-density correlation are close to those in the ground state, and the time-averaged bond density is inverted. Around $ eaF/\hbar\omega $=4 and $ eaF/\hbar\omega $=7.3, the total-energy increment shows a local maximum, but the realized state departs from the typical negative-temperature state at $ eaF/\hbar\omega $=0.9. Compared with the latter, the electrons are transferred between dimers and the charge order is thus weakened. 

The interaction-parameter dependences of the time-averaged correlation functions in the negative-temperature state become similar to those in the ground state if the interactions (except the Peierls electron-lattice coupling) are inverted: the repulsive on-site (nearest-neighbor) interaction is converted to an attractive on-site (nearest-neighbor) interaction, and the relative sign between the charge density ($ n_i - 3/2 $) and the intramolecular displacement is inverted. The energy supplied by the photoexcitation goes to these interactions terms. As for the Peierls electron-lattice coupling, the relative sign between the bond density and the intermolecular displacement is inverted for those field amplitudes ($ eaF/\hbar\omega $=0.5 and $ eaF/\hbar\omega $=1.3), which are different from the field amplitude for the negative-temperature state to appear ($ eaF/\hbar\omega $=0.9). The formation of the negative-temperature state and the inversion of interactions have been discussed by Tsuji {\it et al.}\cite{tsuji_prb12} using the dynamical mean-field theory, but they need half-cycle pulses or asymmetric monocycle pulses in the cases studied. In this paper, we demonstrate that they are realized even by symmetric monocycle pulses. 

\begin{acknowledgment}
The authors are grateful to Y. Tanaka for various discussions. 
This work was supported by Grants-in-Aid for Scientific Research (C) (Grant No. 23540426) and Scientific Research (A) (Grant No. 23244062) from the Ministry of Education, Culture, Sports, Science and Technology of Japan. 
\end{acknowledgment}

\bibliography{66977}

\end{document}